\documentclass[conference]{IEEEtran}
\IEEEoverridecommandlockouts
\usepackage{cite}
\usepackage{amsmath,amssymb,amsfonts}
\usepackage{algorithmic}
\usepackage{graphicx}
\usepackage{textcomp}
\usepackage{xcolor}
\def\BibTeX{{\rm B\kern-.05em{\sc i\kern-.025em b}\kern-.08em
    T\kern-.1667em\lower.7ex\hbox{E}\kern-.125emX}}
\begin{document}

\title{D2D-LSTM based Prediction of the D2D Diffusion Path in Mobile Social Networks\\
}

\author{\IEEEauthorblockN{1\textsuperscript{st} Hao Xu}
\IEEEauthorblockA{\textit{College of Intelligence and Computing} \\
\textit{Tianjin University}\\
Tianjin, China \\
vocal7@tju.edu.cn}}

\maketitle

\begin{abstract}

Recently, how to expand data transmission to reduce cell data and repeated cell transmission has received more and more research attention.. In mobile social networks, content popularity prediction has always been an important part of traffic offloading and expanding data dissemination. However, current mainstream content popularity prediction methods only use the number of downloads and shares or the distribution of user interests, which do not consider important time and geographic location information in mobile social networks, and all of data is from OSN which is not same as MSN. In this work, we propose D2D Long Short-Term Memory (D2D-LSTM), a deep neural network based on LSTM, which is designed to predict a complete D2D diffusion path. Our work is the first attempt in the world to use real data of MSN to predict diffusion path with deep neural networks which conforms to the D2D structure. Compared to linear sequence networks, only learn users' social features without time distribution or GPS distribution and files' content features, our model can predict the propagation path more accurately (up to 85.858\%) and can reach convergence faster (less than 100 steps) because of the neural network that conforms to the D2D structure and combines user social features and files features. Moreover, we can simulate generating a D2D propagation tree. After experiment and comparison, it is found to be very similar to the ground-truth trees. Finally, we define a user prototype refinement that can more accurately describe the propagation sharing habits of a prototype user (including content preferences, time preferences, and geographic location preferences), and experimentally validate the predictions when the user prototype is added to 1000 classes, it is almost identical to the 50 categories.
\end{abstract}


\section{Introduction}

As the number of mobile network multimedia services increases, mobile users tend to download files to local mobile devices, which poses challenges for mobile network operators (MNOs) front-haul and back-haul networks\cite{cisco_index_mobile_datatraffic}. 
In this case, the traffic explosion problem often occurs in some areas where the cellular network capacity is limited but the user demand is large. Due to the lack of MNOs infrastructure network facilities and limited wireless link capacity, it is difficult for them to effectively handle the skyrocketing traffic load.

Cha et al. \cite {cisco_index_mobile_datatraffic} \cite {cha_Youtube_ieverybody_IMC07} suggests that users are more likely to download content files of the same popularity because the top 10\% of YouTube videos account for around 80\% of all views. Network resources are wasted because of these repeated traffic.

An effective way is to encourage subscribers to get popular content files (APPs) from devices around them. It can be performed by caching and transmitting popular multimedia content files via geographically available device-to-device (D2D) communication (e.g., Wi-Fi Direct, Bluetooth, etc.), which greatly reduces repeated downloads. Note that we consider various types of device-to-device (D2D) direct communication techniques between devices, including the most recent D2D communication studied in \cite{D2D_3GPP_LTE}, the 3GPP LTE cellular network.

There are many models in terms of content dissemination. In particular, the spread of the virus (also known as "word of mouth" diffusion) \cite{Bak_role_of_social_networks} \cite{Kim_tractable_models} \cite{Liben_link_prediction}  \cite{Rod_word_of_mouth}has proven to be an important mechanism for promoting new ideas, technologies, content or products. Unlike mass broadcast counterparts that transmit information to a large number of users directly, in the spread of viruses, many individuals participate in the dissemination of information in a chain structure.Whether our goal is to cache duplicate files or content dissemination research, content popularity prediction is a very important part, which can directly determine the quality of the cache and the accuracy of content dissemination.

\begin{figure*}[!!htbp]
\centerline{\includegraphics[width=18cm]{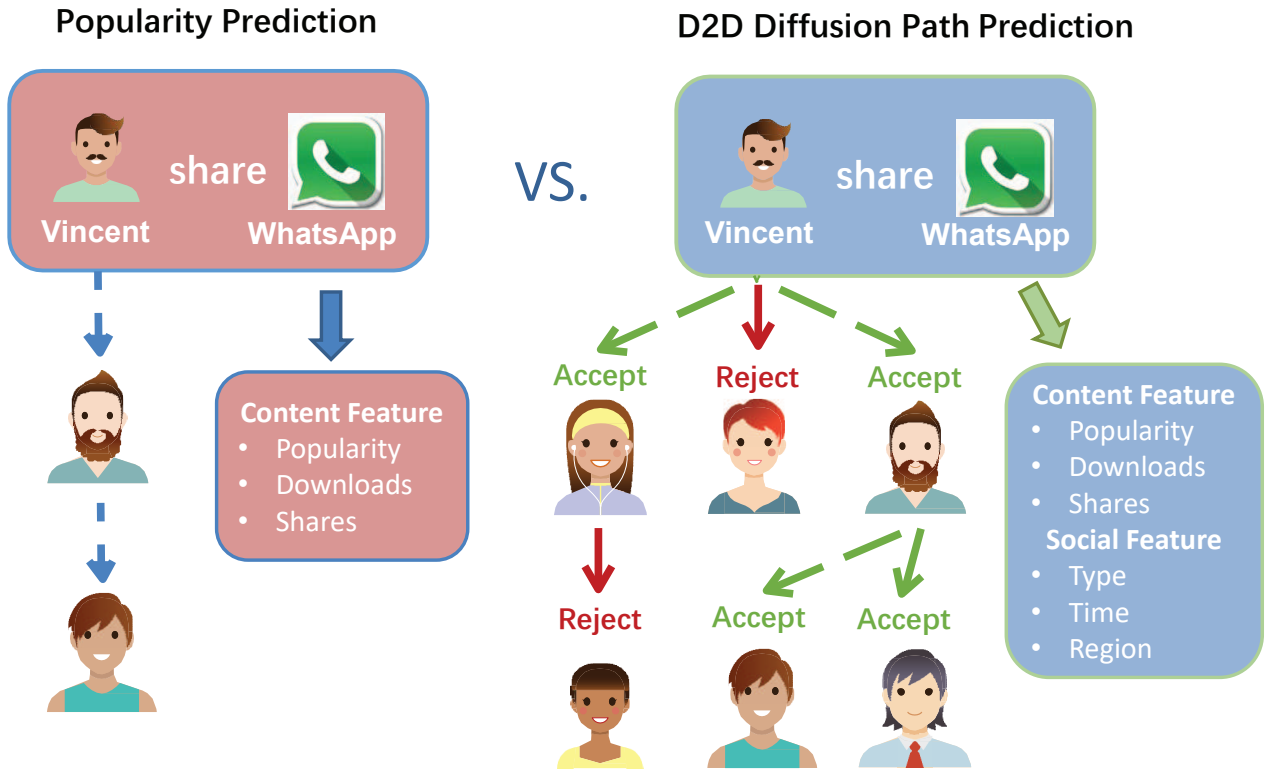}}
\caption{(Left): Previous work on popularity predictions predicts whether a certain type of APP will become popular, as measured by a single metric (such as the total number of users who downloaded or shared the APP). (Right): Our work on diffusion prediction predicts the entire propagation path of the APP. Specifically, our approach recursively predicts which users the user will share, the social characteristics of the user, and the previous sharing history to share/re-share the APP.}
\label{content_popularity_vs_d2d_diffusion}
\end{figure*}

In this article, we are interested in how file content (APP type) is spread through mobile social networks. The reason why we choose APP data analysis is that APP is more and more able to reflect the multi-dimensional characteristics of users' personality, interest, social and so on in MSN because of its rich functions and user-friendly design, which is very beneficial to our research (e.g., WhatsApp, Wechat, FIFA, Youtube). 

In the existing work, such as \cite{wang_data_measure}, this is the multi-dimensional data measurement, analysis and prediction based on Xender data, but the prediction part does not reflect the chain or tree architecture of the data. We expect to use the complete model of data propagation to train, ,valid and forecast rather than a single data record. Moreover, these existing works \cite{aga_clooective_wisdom} \cite{kho_image_popular} \cite{li_online_social} mainly focus on a single feature, such as the content type feature (e.g., the number of sharing times, the number of downloading times) (see Fig.~\ref{content_popularity_vs_d2d_diffusion} left), We hope that the entire structure of the content propagation path can be modeled to bring the model closer to the real D2D propagation process, thereby increasing the prediction accuracy (see Fig.~\ref{content_popularity_vs_d2d_diffusion} right). Furthermore, the existing work \cite{hu_share_prediction} are concerned with propagation predictions on OSN, which is different from MSN. MSN propagation has distance limitations and has an important factor in geographic location. Thus, we propose D2D-LSTM, a real D2D data based deep recurrent network that recursively predicts each level of the APP content diffusion path utilizing the content features (APP type) and all users' social features (e.g., interest distribution, time distribution and region distribution) in the diffusion path, including previous and present. As we know, users don't just have an interest in apps. For example, users who like sports also like social apps. Moreover, their classification of user prototypes is not diversified enough to accurately reflect the user's true preferences.

In our model, we use the APP type and the user's social properties and focus on all the previous records to accurately predict the sharing propagation path. Therefore, because our historical factors are taken into account in our model, our predictions are also determined by historical features and current features, so that we can predict a complete propagation path better than ever before. From this we can know that the size of a file's ability to spread is not only related to the category of the file itself, but also to the social attributes of the users who spread it (interest distribution, time distribution, geographical distribution). Therefore, we designed our D2D-LSTM to capture the history of all previously shared APP files and the history of the user's social features, as well as current properties, in order to predict their future propagation paths. In addition, in order to adapt our prediction model to new users, we group prototype users based on the social capabilities of individual users. To the best of our knowledge, our work is the first attempt to generate a real D2D data based complete diffusion path for APP categories through mobile social networks. Our results show superior diffusion path prediction performance compared to shared predictions that lack historical memory or rely solely on a single information feature, even online social networks. we define a user prototype refinement that can more accurately describe the propagation sharing habits of a prototype user (including content preferences, time preferences, and geographic location preferences), and experimentally validate the predictions when the user prototype is added to 1000 classes. Not only that, although we use Xender data to evaluate our models, we design them to be generic and also apply to other D2D mobile social networks.

\section{Related Work}

In this section, we discuss related work in following three aspects:
Features and Propagation models.

\subsection{Mobile Social Networks Based D2D sharing path}

Recently, more and more researches have focused on the use of D2D communication for MSN content sharing \cite{Rod_word_of_mouth} \cite{han_mobile_data_offloading} \cite{xiang_model_relationship}. Many mobile applications, such as Apple's Airdrop \cite{airdrop_apple} and Xender have also been promoted. Most of these researches concentrate on the D2D epidemic content dissemination in MSNs recently with the purpose of offloading reduplicate traffic loads. Zhang et al. \cite{zhang_performance_cn} and Li et al. \cite{li_energy_tvt} propose differentiation-based models to measure the performance of popular content sharing in MSNs. Nonetheless, they rarely perform over a large-scale user base but a group consists of small amounts of people.

Not only is the prediction of shared content on mobile social networks, but there are also a lot of efforts in online social networks that have explored predictions including images \cite{cheng_predict_cascades} \cite{gelli_image_predict}, video \cite{li_popular_prediction}, tweets \cite{hong_predict_twitter} \cite{kup_predict_tweet} and other methods of internal popularity, these models also have function that combine content functions with user social functions. 

In fact, the temporal and spatial constraints make it more complex to analyze D2D sharing in offline MSNs than information sharing in OSNs. There are studies like \cite{rodrigues_on_imc} \cite{ke_understanding_www_homophily} \cite{wittie_exploit_conext} focusing on the homophily and locality features of users in both MSNs and SNSs, facilitating the D2D content dissemination \cite{wang_data_measure}. 
Unlike these existing efforts that mobile social networks mainly focus on single sharing focus primarily on predicting the popularity score of content (e.g., the number of shares), our goal is to predict the entire content propagation path through the D2D mobile social network, which is a more realistic situation, naturally It is also a more challenging task. Furthermore, unlike the memory-based deep recurrent network in \cite{hu_share_prediction} to predict the image propagation path in the OSNs, our D2D-LSTM pays more attention to the influence of geographical location on the propagation path of the mobile social network, making the prediction more consistent to the model of mobile social networks  communication.  Moreover, their classification of user prototypes is not diversified enough to accurately reflect the user's true preferences.

We first need to predict the number of user types that share the APP, and further predict the prototype category characteristics of each sharer in the diffusion sequence, so that we can more accurately predict the entire APP propagation process. Then we we define a refined user prototype classification that can more accurately describe the propagation sharing habits of a prototype user (including content preferences, time preferences, and geographic location preferences).

\subsection{Propagation Models}

We use our D2D-LSTM architecture to predict the D2D mobile social network propagation tree and correlate it with Tree-LSTM \cite{tai_improved_sementiic} \cite{zhang_top_down} in Natural Language Processing (NLP), but it is worth noting that our model structure is Unlike the standard linear chain-LSTM, Tree-LSTM uses a bottom-up approach to predict the next hidden state from hidden vectors from multiple children in the same layer, and has achieved significant results in natural language processing such as sentiment analysis. Unfortunately, unlike the D2D mobile social network propagation model, starting from the root node, according to the multi-child tree structure, the tree-lstm is the bottom-to-child node from the bottom up to the parent node's emotion prediction, and finally Predict the emotions of the root node \cite{zhang_top_down}. Therefore, our D2D-LSTM, which is designed to conform to the propagation mode of the D2D mobile social network, generates a content diffusion path for the mobile social network tree prediction in a top-down manner. To the best of our knowledge, our work is the first attempt to generate a complete diffusion path for an APP through a D2D mobile social network.

\section{Approach}

\subsection{Xender Data}

Xender is a mobile application (APP) based on D2D content delivery which is popular worldwide rely on providing users with the ability to share on various mobile platforms (e.g., Android, iOS and Windows) without using the 3G / 4G cellular network infrastructure. The D2D connection in Xender does not use mobile data, mainly based on Wi-Fi network sharing, and also supports WiFi Direct and Bluetooth, thus the transmission is free. 

\begin{figure}[!ht]
\centerline{\includegraphics[width=9cm]{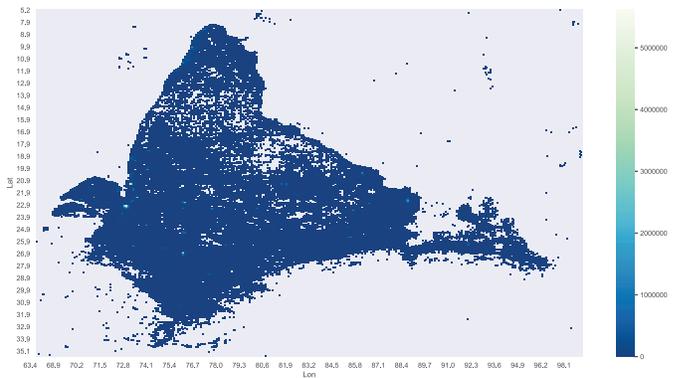}}
\caption{India GPS heatMap whitch is from 1st February,2016 to 16th February,2016.}
\label{gps_heatmap}
\end{figure}

Xender's daily content delivery is about 110 million, with about 9 million users per day and 100 million monthly active users. From February 1, 2016 to February 28, 2016, we captured the Xender trail for a month. 

\begin{figure}[ht]
\centerline{\includegraphics[width=9cm]{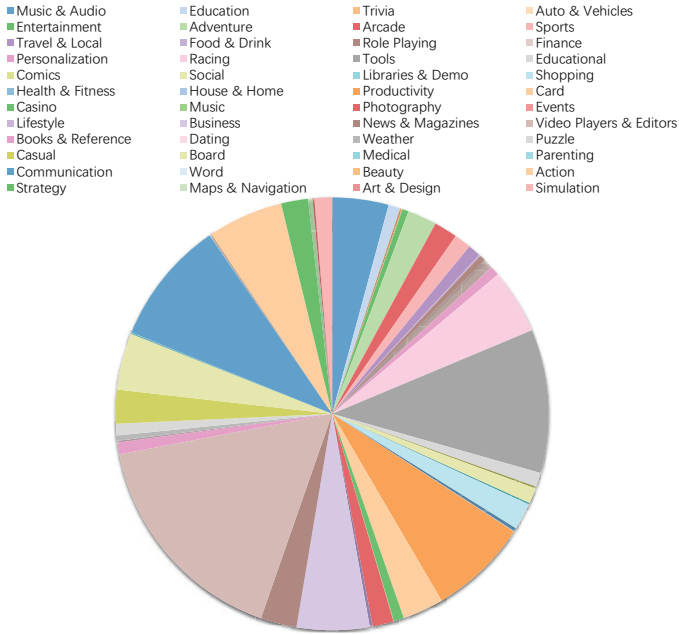}}
\caption{48 types of application.}
\label{app_type}
\end{figure}

\begin{figure*}[!!hbt]
\centerline{\includegraphics[width=17cm,height=10cm]{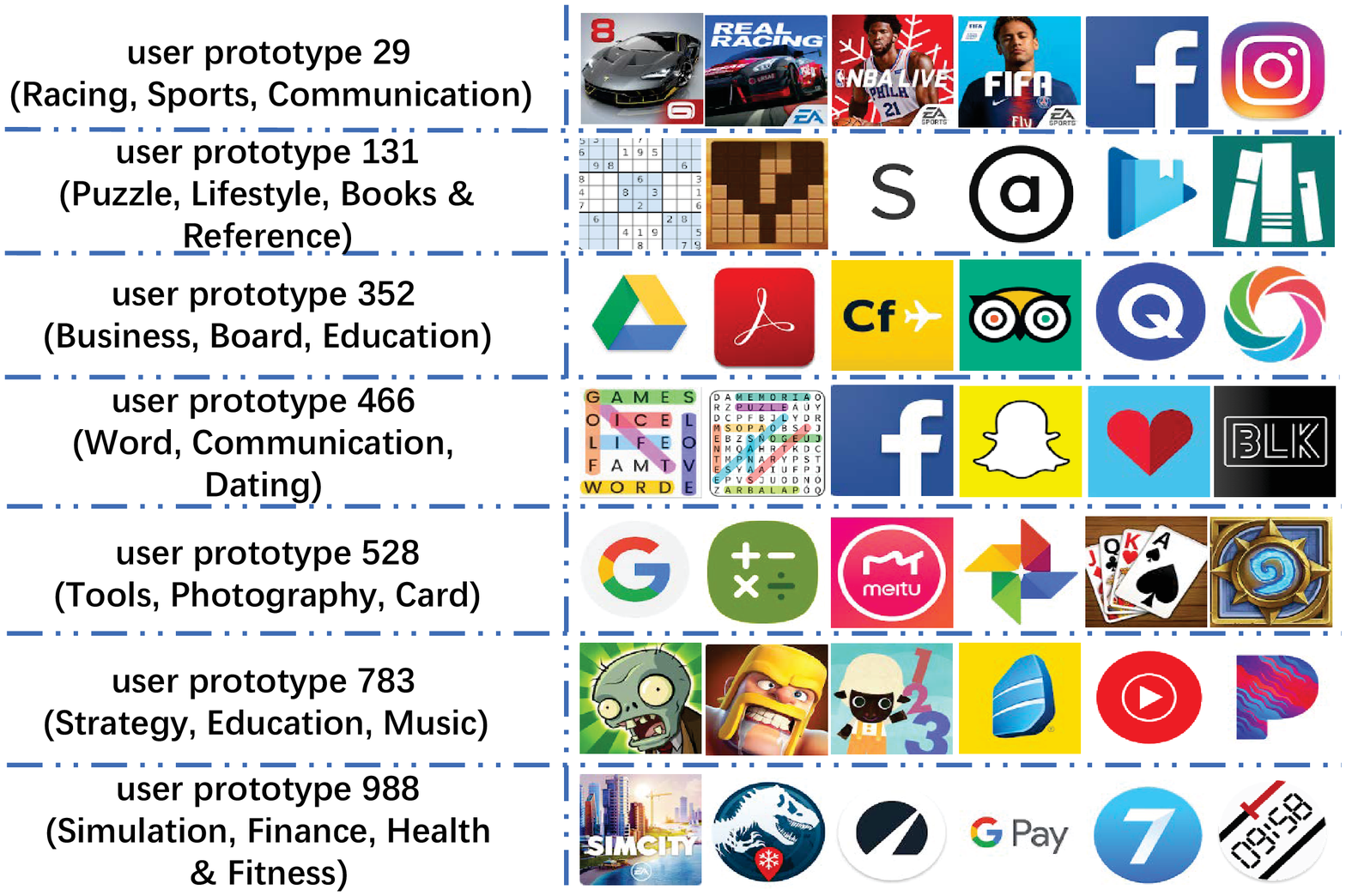}}
\caption{Examples of a prototype user. The left panel shows the first three APP categories of the APP that are most commonly shared by the prototype user, while the right panel displays the specific APPs that are shared by each user mapped to the prototype.}
\label{prototype}
\end{figure*}

In this article, we focus on predicting the sharing path in Indian users'  share action based on D2D mobile social networks because we found that about 70\% of users are from India (2 weeks GPS heatmap is as shown in the Fig.~\ref{gps_heatmap}), which may be due to their limited cellular network services. After cleaning unnecessary and invalid data (e.g., incomplete format, zero size file, user with invalid GPS), there are a total of 30,485,335 users, of which 4,434,440,043 transfers transmitted 16,785,175 content files. 

We use the APP categories of Google Play \cite{google} to categorize APPs that are distributed by Xender. There are 48 categories including photography, social, and tools. The APP type distribution is as shown in the Fig.~\ref{app_type}. 

From the Fig.~\ref{app_type} we can see that the apps that are spread in the D2D mobile social network are different in type, and the number of times of being spread is also different. Video Players \& Editors is the most popular type of APP, accounting for 17\%. The Tools category followed, with 11\% of users choosing to share such apps, while Art \& Design had the least share, less than 1\%.

\subsection{Data Preprocess}

For our next work (building a user prototype), we need to capture the characteristics of the user's social characteristics, including the distribution of preferences for the type of APP, the distribution of sharing tendency, the habit distribution for transmission time, and the geographical distribution of the regions that are often in place.

\begin{itemize}
\item 48 types of application is used by us, based on the classification of apps in google play.
\item 2 dimensions vector which contains the number of shares and receives.
\item 24 time periods are used to reflect the time habits of Xender users sharing apps. Each session lasts for one hour.
\item 1000 geographical locations are used to reflect the geographical distribution of users' shared apps. Due to the accuracy and dispersion of geographic location GPS, we use k-means to cluster GPS. We set k=1000 clusters so that we can objectively respond to the geographical distribution of a user's entire transmission history.
\end{itemize}

\subsection{Mapping Xender Users to D2D User Prototype}

\begin{figure*}[!!hbt]
\centerline{\includegraphics[width=18cm]{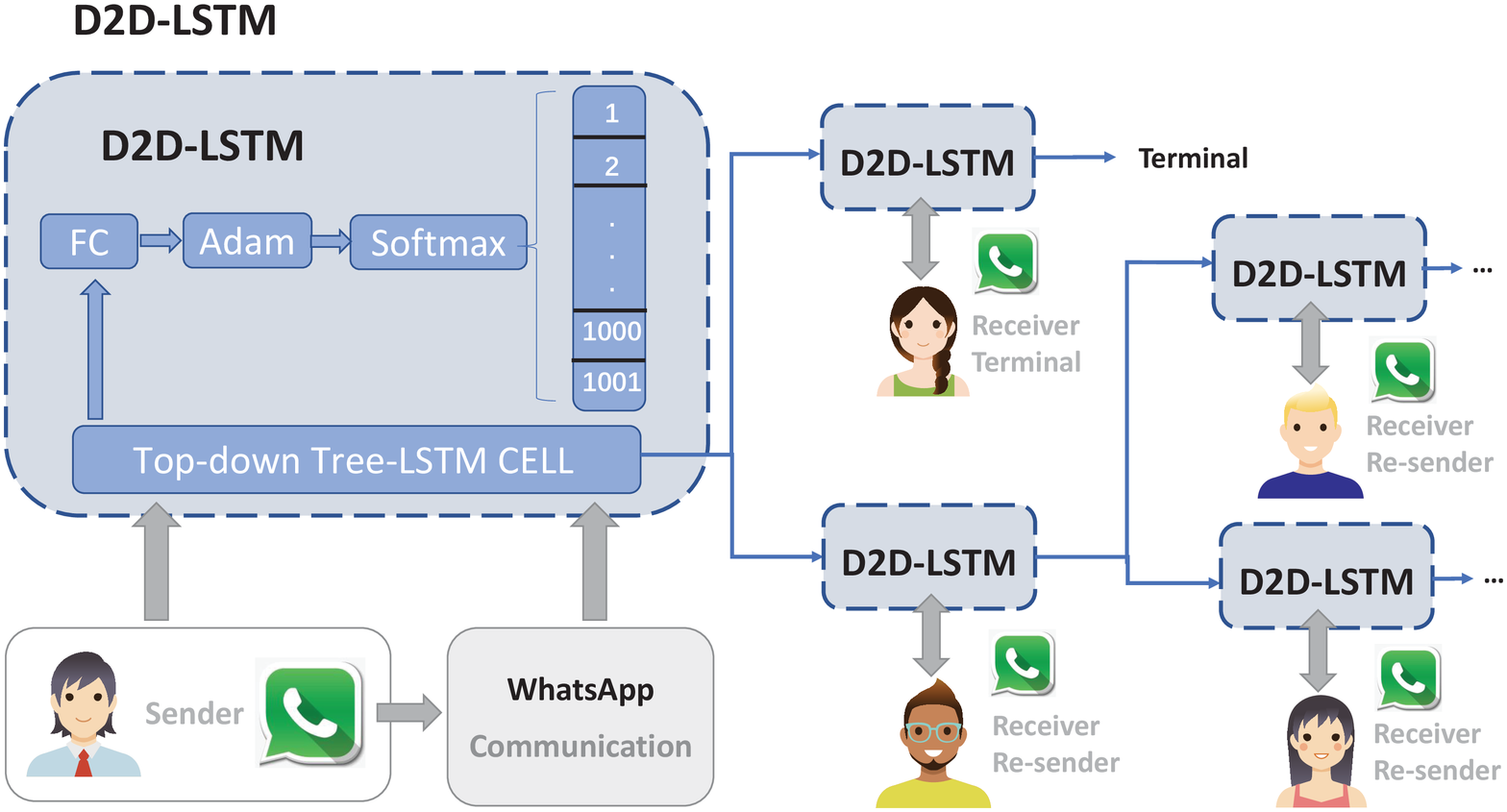}}
\caption{The architecture of the D2D-LSTM model. The content feature of the shared APP is fed as an initial memory $c_0$ to the D2D-LSTM network. In each step, the D2D LSTM unit takes the current user's social features (type distribution, sharing tendency, time distribution, GPS distribution) as input. The hidden vector output of D2D-LSTM unit $h$ passes through the FC layer and Adam optimizer, followed by a softmax function to obtain 1001 prototype probabilities (1000 prototype users and 1 terminal prototype). The D2D-LSTM unit in the dashed rectangle repeats at successive tree nodes, the only difference between the root node and the child node is that the child node obtains memory content $c$ and output $h$ from its parent node.}
\label{d2d_lstm_constructure}
\end{figure*} 

Based on our goal is to predict the complete propagation path of the APP in the D2D mobile social network. We first need to explain how we map individual users to a set of prototype users, so that our proposed model can be extended to users outside the experimental data to increase the universality of the model.

The APP's diffusion path encodes the user sequence that shares it. The set of all possible diffusion paths depends on the users in the mobile social network as well as the network structure (e.g., the connections between users). Because we want our model to be able to extend the sharing relationship between users and users outside the experimental dataset to increase the universality of the model. So we need to generalize a set of user prototypes that contain many similar users to capture the general shared features that map to each user (such as the APP categories they like to spread, the time of sharing behavior and the area in which they are often shared). In this way, we can ignore the nuances of real users during training and testing, and only train the same prototype users. Not only that, but we can combine similar individual users into a single prototype user to expand the training data to prevent a user's training data from being insufficient, which can produce a better prediction model. This has been verified in previous work \cite{wang_data_measure} that size of the spread tree of some APPs is limited, and the depth is about 4 layers, so the prototype user is necessary.

Specifically, due to the variety of app categories, times, and geographic locations in the Xender dataset, we use k-means to describe the user's social characteristics (described in more detail below, and we will discuss the effect of experimental prediction accuracy with different number of clusters in later chapters). Clustering is k = 1000 prototype groups. We set the number of prototypes to 1000 to cover a broad range of users. According to the experimental results, We find that our model can adapt to a wide range of k values. We can map users outside the experimental cluster to the prototype (e.g., the cluster center) reveals the average social feature of all its cluster members, making their original social features most similar based on their Euclidean distance. Fig.~\ref{prototype} shows an example of a prototype user. The left panel shows the first three APP categories of the most frequently shared APP which is from Xender dataset of sharing trace, while the right panel displays specific APPs shared by each user mapped to the prototype. For example, prototype user 528 prefers to share communication and social, while prototype user 763 likes to share actions and puzzles.

\subsection{Network Architecture}

The architecture for modeling D2D diffusion through mobile social networks is shown in Fig.~\ref{d2d_lstm_constructure}. The input of our model is the APP content attribute $C$ shared by the user and the social attribute $S$ of the user. The output is a serialized propagation of the entire diffusion tree $\{P1, P2, ..., Pn\}$. Specifically, the initial diffusion path is conditioned on the input APP content attribute $C$ and the social attribute $S$ of the root user. Then, in each subsequent time step, our model records the social characteristics of the parent node user and the historical state of the social features of the current point, and outputs the re-shared target user prototype, including 1000 types of prototype users and one terminal prototype with a total of 1001 categories. The model stores and updates a memory that records the characteristics of the user in the diffusion path so far. The process ends when the next predicted user is a "terminal prototype". For the APP content attribute, we divided it into 48 categories according to Google Play's classification, which has been mentioned before. For the each user' social feature, we calculate the 48 types of APP types shared by users, the 2 dimension vector of sharing and accepting APP, the 24 hours distribution, and the GPS distribution of 1000 classes, for a total of 1074 dimensional vectors.

\subsection{D2D-LSTM}

\begin{figure*}[!!hbt]
\centerline{\includegraphics[width=18cm]{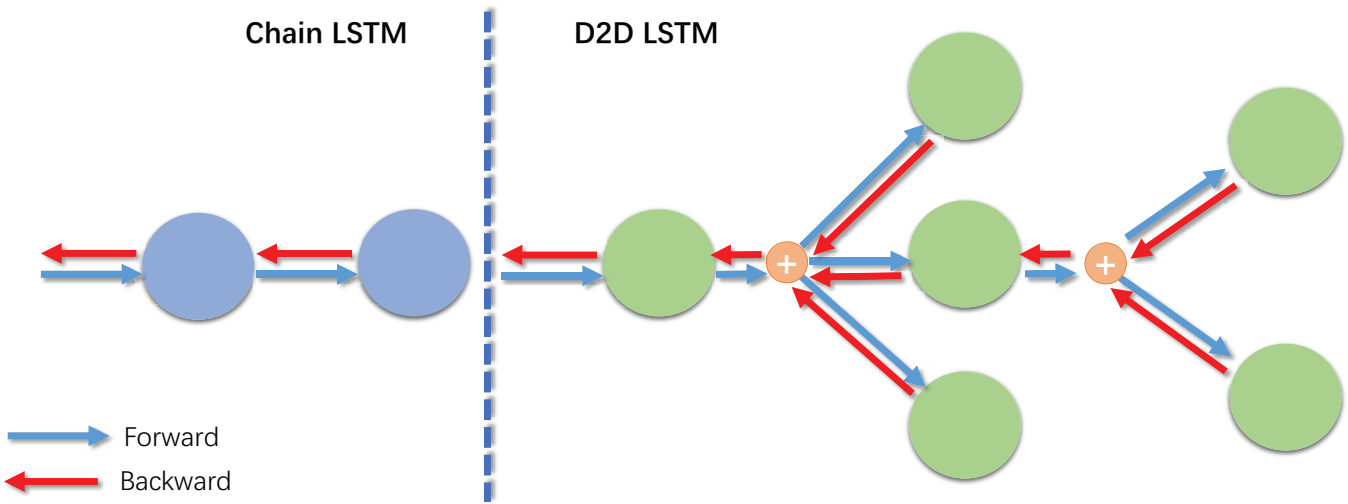}}
\caption{The blue and red lines indicate forward and backward propagation, respectively. (Left): LSTM network with chain structure which nodes are arranged according to the time step $t$. (Right): D2D-LSTM network which nodes are scheduled based on share relationships. The sum of the child nodes is passed to the parent node during the backward propagation period.}
\label{chain_vs_d2d}
\end{figure*}

Whether any APP will be propagated further at any layer along the propagation path will depend not only on the social attributes of the current user, but also on the social impact of his parent node which contains the history of the diffusion path taken so far. This is an inherent recursive problem that is well suited for analyzing sequences D2D diffusion model using recurrent neural networks. Because of the vanishing gradient problem in the long-range dependencies across many diffusion steps, we use the LSTM model \cite{gers_learning_forget} \cite{ho_lstm} instead of the RNN \cite{ben_gradient}. In our example, the APP content feature is combined with the current user's social feature to recursively predict the next user to share (as shown in the right image of Fig.~\ref{content_popularity_vs_d2d_diffusion}). The only difference between tree structure D2D-LSTM network and conventional chained structure LSTM network is the flow path of cell state $c$ and the hidden state $h$ following the tree. In a conventional LSTM, the previous memory cells and hidden states starting at time step $t-1$ are passed to a single next node in time step $t$. However, in our D2D-LSTM network, the hidden state of the previous memory unit and parent node starting at time step $t-1$ can be passed to multiple child nodes in time step $t$, as shown in Fig.~\ref{chain_vs_d2d} This means that during backward propagation, the gradient of the parent node sums the gradients passed by all of its child nodes. Formally, the subscripts $cd$ and $pt$ are used to represent "child" and "parent" respectively. The D2D-LSTM parent-child transfer equation is expressed as follows:

\begin{align}
i_{cd} &= \mathrm{sigmoid}(W^{(i)} x_{cd} + U^{(i)} {h}_{pt} + b^{(i)}) \\
f_{cd} &= \mathrm{sigmoid}(W^{(f)} x_{cd} + U^{(f)} h_{pt} + b^{(f)}) \\
o_{cd} &= \mathrm{sigmoid}(W^{(o)} x_{cd} + U^{(o)} {h}_{pt} + b^{(o)}) \\
u_{cd} &= \tanh(W^{(u)} x_{cd} + U^{(u)} {h}_{cd} + b^{(u)})
\end{align}
\begin{align}
c_{cd} &= i_{cd} \circ u_{cd} + f_{cd} \circ c_{pt} \\
h_{cd} &= o_{cd} \circ \tanh(c_{cd})
\end{align}

Where $W$, $U$ and $b$ are the weight matrices and bias vectors to be optimized, $x_{cd}$ and $i_{cd}$ are the input social feature of the user at the current step and input gate vector which us used for acquiring new information. $◦$and $o_{cd}$ denotes elementwise multiplication and an output gate vector, and $f_{cd}$ denotes the forget gate vector which is used for remembering old information, $c_{cd}$ denotes the memory cell, and $h_{cd}$ denotes the hidden state.

Our D2D-LSTM model consists of the Top-down Tree LSTM unit followed by one FC layer and one softmax layer afterwards to predict one of the k = 1001 prototypical users including 1000 "prototype" class and 1 "terminal" class. To reduce overfitting and increase the nonlinearity of the model, we add one dropout layer after the Top-down Tree-LSTM unit, and one Adam layer and one dropout layer between FC layer and softmax layer. The initial hidden state $h_0$ is simply set to all zero. Then we add the file attribute and the user's social attribute so that each layer of the propagation contains the user's social and file features.

\subsection{Loss Function}

The propagation of files for each layer in the propagation tree can be seen as a multi-classification problem, in which the user social attributes of the current layer and the file attributes are used to predict the re-shared user prototypes of the next layer. In addition, we can classify users from $0$ to $n$ prototypes, finally we predict the distribution of $n+1$ (terminal class) users in the next layer. So we use cross entropy loss to characterize cost values and Adam function to optimize our model:

\begin{align}
loss(x, prototype) &= -log(e^{x_{prototype}} / \sum\nolimits_{k}^{N}e^{x_{k}}) \nonumber \\
&= -x_{prototype} + log(\sum\nolimits_{k}^{N}e^{x_{k}})
\end{align}

where $x_prototype$ indexes the classes (prototype users and terminal class), $k$ represents the $k$th value of the predicted probability weight $e^{x}$, then we increase forecast accuracy by keeping the loss down.

\section{Experiments}

In this section, we evaluate the accuracy of our model’s APP diffusion path prediction, and compare to several baselines that either use only content or social features, or lack memory. We also show qualitative examples of our model’s generated diffusion trees. 

\subsection{Model Details}
We train our D2D-LSTM model with an initial learning rate of 0.1, and then lower the learning rate to 0.01 when validation loss stops decreasing. We train both our model and all baselines within 40-80 epochs depending on the model until the models are full convergence. We use back propagation over time to optimize the weight of the model, then we divide the social feature by the maximum value to normalize it to [0,1] and set the content feature value to 1 to correspond to the social feature.

\subsection{Baselines}
We compare the D2D-LSTM model to several baselines: FC model with content and social features: The baseline consists of three fully connected (FC) layers. An Adam layer and a dropout layer follow each of the first two FC layers. The last FC layer is followed by a softmax layer that converts the output weights into probabilities. This baseline is chosen to investigate the importance of memory for content diffusion path prediction. D2D-LSTM only has content capabilities: it has the same architecture as our D2D-LSTM model, but only has content capabilities. D2D-LSTM with content capabilities and some social features: The same architecture as the D2D-LSTM model, but with content features and some social features such as GPS distribution, time distribution and content distribution. For both baselines, the initial memory c0 is also set to all zeros. These two baselines measure the effect of social features on content diffusion path prediction. Note that because of the bottom-up nature of Tree-LSTM \cite{zhu_lstm} \cite{Tai_tree_lstm} we can't compare it to our problem because we focus on top-down propagation tree prediction/generation.

\subsection{Prediction Loss and Accuracy}
From Table \ref{tab1} We first evaluate the APP diffusion prediction by comparing our predicted users to the ground-truth users of each node in the tree. We see that adding social features produces better predictions than individual APP content features. In social characteristics, the importance of GPS information for prediction accuracy is significantly higher than other features (e.g., time characteristics, etc.), and the accuracy of prediction after Time feature is increased by 2.155\%, and the accuracy of prediction after adding GPS features is improved. 7.054\%, this is because the user's offline communication is closely related to the geographical location. Second, our Diffusion-LSTM model performs better than the FC model, indicating that the published content share history is stored in memory to predict the key role of the next re-share user. 

\begin{figure}[!!hbt]
\centerline{\includegraphics[width=9cm]{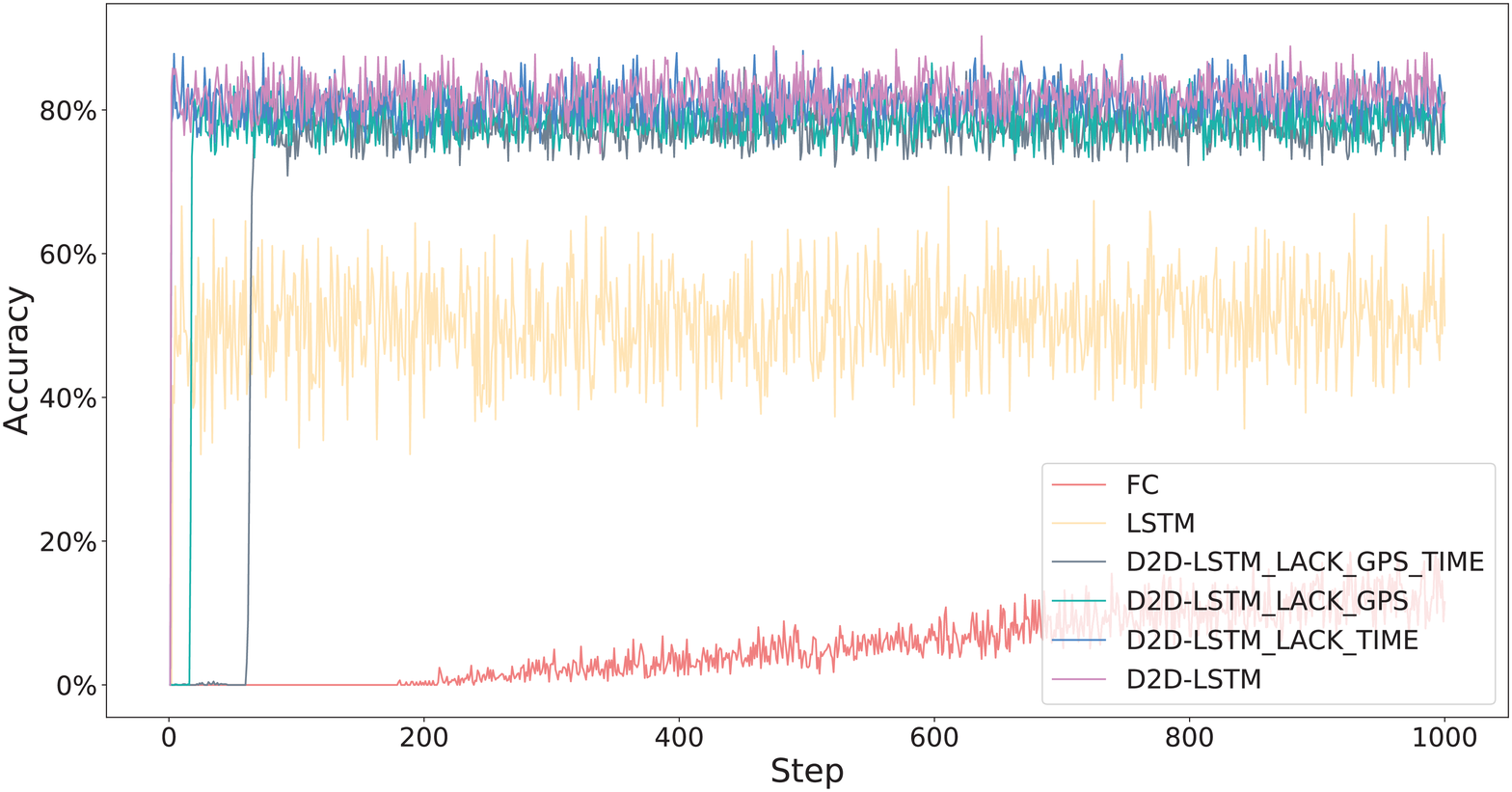}}
\caption{}
\label{baselines_accuracy}
\end{figure} 

\begin{figure}[!!hbt]
\centerline{\includegraphics[width=9cm]{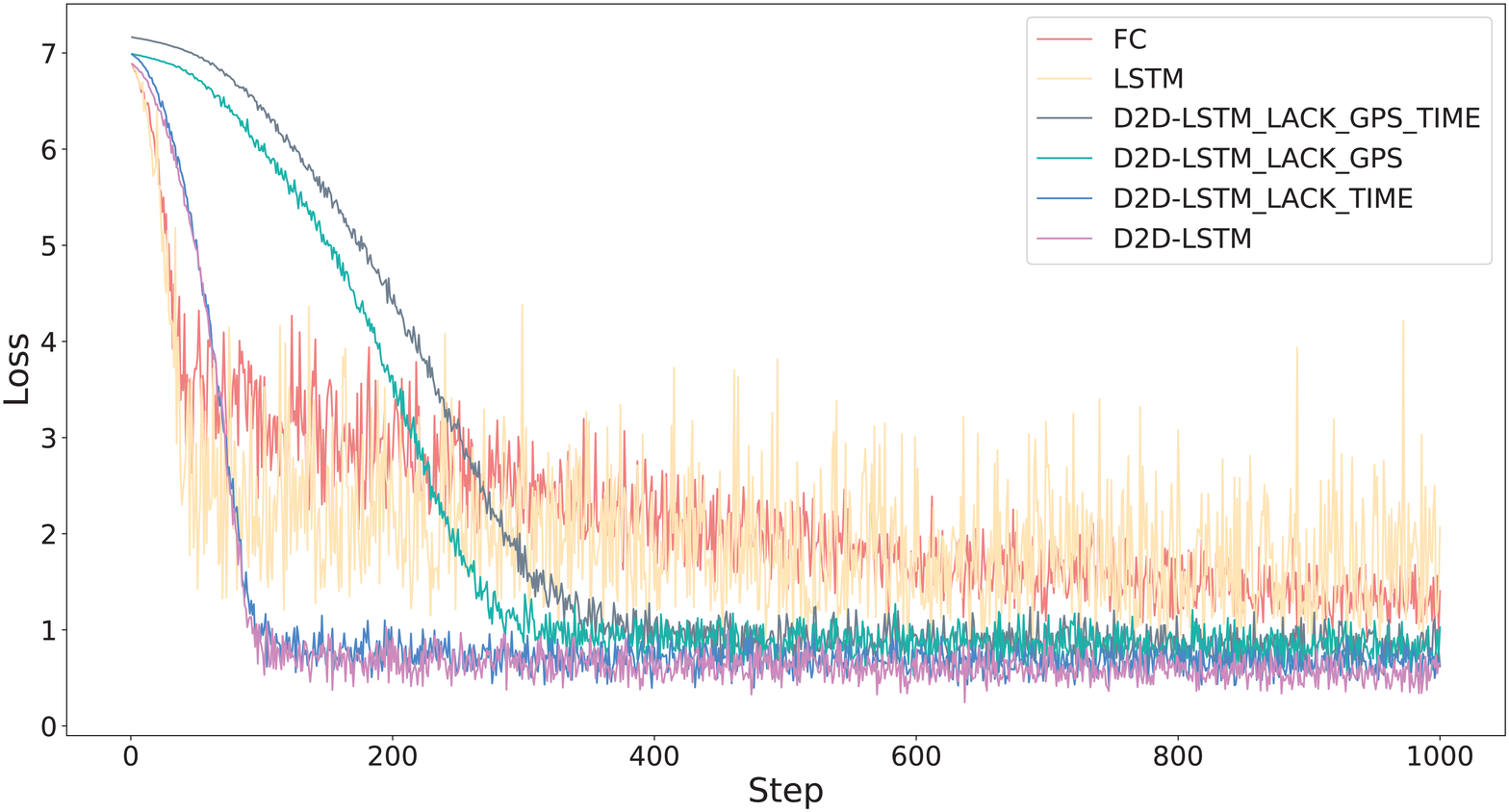}}
\caption{}
\label{baselines_loss}
\end{figure}

Finally, as can be seen from Fig.~\ref{baselines_accuracy}), the combination of social functions and APP content functions can produce the best results. The test accuracy of our D2D-LSTM model is up to 83.787\%. As shown in Fig.~\ref{baselines_loss}, the loss convergence rate of the D2D-LSTM model with all social features and content features is also the fastest. It can be understood that when the propagation characteristics of the model learning increase the geographical location information and time distribution, The convergence speed will be faster (at least 100 steps), and the loss can also reach a better value (optimal to 0.6307).

\begin{table*}[htbp]
\caption{}
\begin{center}
\begin{tabular}{|c|c|c|c|c|c|c|c|c|c|c|c|}
\hline
Model & 
Content Feature & 
\multicolumn{4}{c|}{Social Feature} &
\multicolumn{2}{c|}{Accuracy} & 
\multicolumn{2}{c|}{Loss} & 
Memory & 
Convergence Step \\
\cline{3-10}
 & & Type & Share & Time & Region& Train & Test & Train & Test & & \\
\hline
FC & Yes & Yes & Yes & Yes & Yes & 19.856\% & 17.803\% & 0.9957 & 1.3601 & No & 900 \\
\hline
LSTM & Yes & Yes & Yes & Yes & Yes & 56.355\% & 52.012\% & 1.4371 & 1.7468 & Yes & 200 \\
\hline
D2D-LSTM & Yes & Yes & Yes & No & No & 78.758\% & 75.431\% & 0.5619 & 1.2047 & Yes & 400 \\
\hline
D2D-LSTM & Yes & Yes & Yes & Yes & No & 79.687\% & 75.937\% & 0.5825 & 1.1623 & Yes & 320 \\
\hline
D2D-LSTM & Yes & Yes & Yes & No & Yes & 82.401\% & 80.836\% & 0.4828 & 0.9190 & Yes & 110 \\
\hline
D2D-LSTM & Yes & Yes & Yes & Yes & Yes & 83.787\% & 82.991\% & 0.4803 & 0.6307 & Yes & 100 \\
\hline
\end{tabular}
\label{tab1}
\end{center}
\end{table*}

\subsection{Refinement Prediction}

\begin{figure}[!!hbt]
\centerline{\includegraphics[width=9cm]{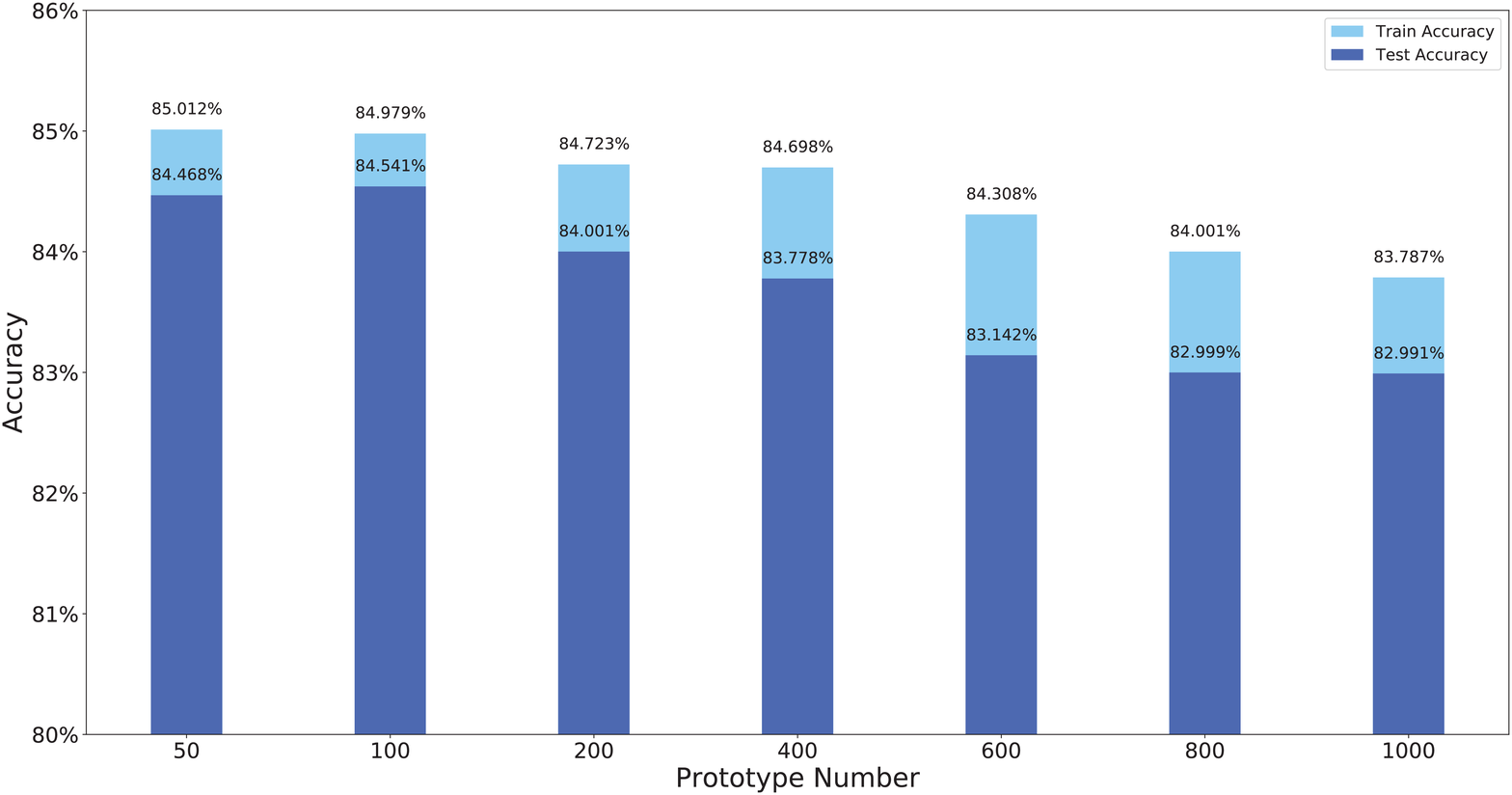}}
\caption{}
\label{change_num}
\end{figure}

In order to test the user's prototype clustering number is different, that is, the degree of refinement of the classification changes, the prediction accuracy of our model, we set the number of user prototypes to 50, 100, 200, 400, 600, 800, 1000, as seen in Fig. \ref{change_num}, As the number of user prototypes increased, the test accuracy dropped from 84.468\% to 82.991\%. Although the accuracy rate has dropped slightly, it still shows that our model can support more refined user prototyping and prediction. Next we continue to increase the number of prototypes to 2000 and 3000 to test the impact of the larger number of prototypes (more granular user classification) on the model.

\subsection{Tree Generation}

\begin{figure}[!!!hbt]
\centerline{\includegraphics[width=9cm]{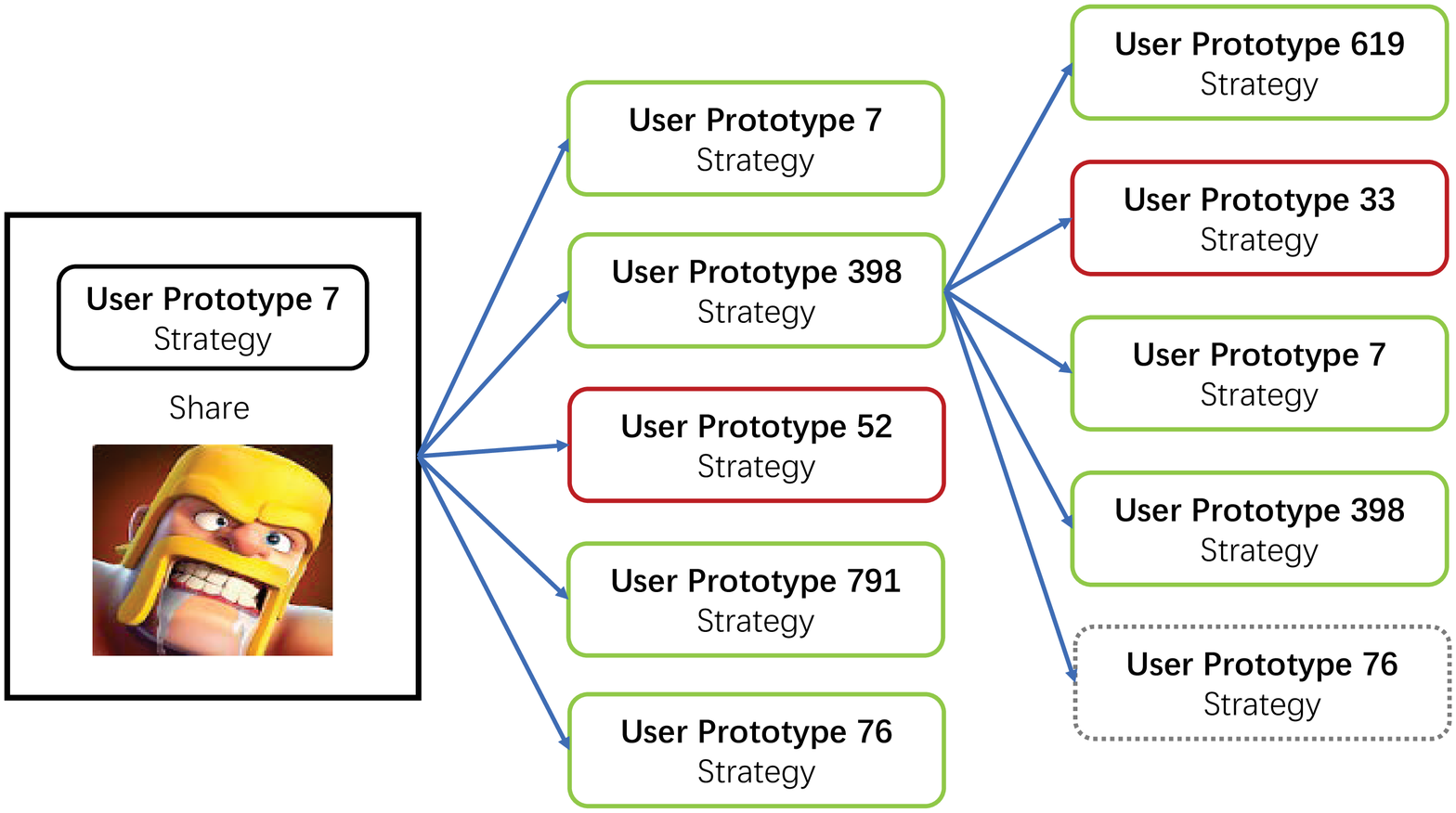}}
\caption{}
\label{prediction_result}
\end{figure}

We then use our trained D2D-LSTM model to generate the propagation tree directly. We compare the generated tree to the ground truth tree, giving the same starting user and shared APP content. In Fig. \ref{prediction_result}, we show the results of a propagation tree generated when a prototype user shares a class of APP. The green solid line represents the prediction accuracy, the red solid line represents the prediction error, and the gray dotted line represents the presence of real data without prediction.

\section{Conclusion}
We propose a deep recurrent network for generating a diffusion path for an APP through a mobile social network. By tracking the spread history of the app, our D2D-LSTM is superior to alternative baselines that lack memory or use only APP content or social features. We further demonstrate the ability of our model to generate meaningful propagation trees. Our model can be applied to a wide range of offline real data sets with versatility and universality. By learning real data, a complete MSN propagation tree is predicted. Finally, our model can also be applied to more refined user prototypes, meaning that our predictions can be more precise and universal.

\section*{Acknowledgment}

The preferred spelling of the word ``acknowledgment'' in America is without 
an ``e'' after the ``g''. Avoid the stilted expression ``one of us (R. B. 
G.) thanks $\ldots$''. Instead, try ``R. B. G. thanks$\ldots$''. Put sponsor 
acknowledgments in the unnumbered footnote on the first page.

\end{document}